\documentclass[twocolumn,showpacs,amsmath,amssymb,superscriptaddress,aps,prc]{revtex4-2}
\usepackage{epsfig}
\usepackage{amsmath}
\newcommand{\cara}{\mbox{$^{16}{\rm C}$}}
\newcommand{\carb}{\mbox{$^{17}{\rm C}$}}
\newcommand{\carc}{\mbox{$^{18}{\rm C}$}}
\newcommand{\cdp}{\mbox{$^{16}{\rm C}(d,p)^{17}$C}}
\newcommand{\cpd}{\mbox{$^{17}{\rm C}(p,d)^{16}$C}}
\newcommand{\card}{\mbox{$^{16}{\rm C}+d$}}
\newcommand{\carp}{\mbox{$^{17}{\rm C}+p$}}
\newcommand{\carn}{\mbox{$^{16}{\rm C}+n$}}
\newcommand{\carzn}{\mbox{$^{16}{\rm C}(0^+)+n$}}
\newcommand{\cartn}{\mbox{$^{16}{\rm C}(2^+)+n$}}
\newcommand{\emax}{\mbox{$E_{\rm max}$}}
\newcommand{\lmax}{\mbox{$\ell_{\rm max}$}}
\newcommand{\elab}{\mbox{$E_{\rm lab}$}}
\newcommand{\ecm}{\mbox{$E_{\rm c.m.}$}}
\newcommand{\rms}{\mbox{$\sqrt{<r^2>}$}}
\newcommand{\etal}{\sl et\ al.}

\begin{document}
\title{Analysis of the $\cdp$ reaction from microscopic $\carb$ wave functions}
\author{Le Hoang Chien}
\email{chienlhphys@gmail.com}
\affiliation{Physique Nucl\'eaire Th\'eorique et Physique Math\'ematique, C.P. 229, Universit\'e Libre de Bruxelles (ULB), B 1050 Brussels, Belgium}
\affiliation{Department of Nuclear Physics, Faculty of Physics and Engineering Physics, University of Science, Ho Chi Minh City, Vietnam}
\affiliation{Vietnam National University, Ho Chi Minh City, Vietnam}
\author{P. Descouvemont}
\email{pierre.descouvemont@ulb.be}
\affiliation{Physique Nucl\'eaire Th\'eorique et Physique Math\'ematique, C.P. 229, Universit\'e Libre de Bruxelles (ULB), B 1050 Brussels, Belgium}
\date{\today}
\begin{abstract}
We present a semi-microscopic study of the $\cdp$ transfer reaction.  The $\carb$ overlap integrals
and spectroscopic factors are obtained from a microscopic cluster model, involving many $\carn$
configurations. This microscopic model provides a fair description of the $\carb$ bound-state energies. 
The $\card$ scattering wave functions are defined in the CDCC method, where the deuteron breakup is 
simulated by pseudostates. The transfer cross sections are in good agreement with recent data.
We confirm the $\cartn$ structure of the ground state,  and show that deuteron breakup effects have a 
significant influence on the cross sections. We study the $\cpd$ reverse reaction
and suggest that the cross section to the $2^+$ state should be large.  
A measurement of the ground-state cross section would provide a strong test of the microscopic 
wave functions. 
\end{abstract}
\maketitle
	
\section{Introduction} 
\label{sec1}
The physics of exotic nuclear is one of the main interests in modern nuclear physics \cite{TSK13,OGS20}.
Exotic nuclei are located near the drip lines and therefore present a low breakup threshold, 
and a small number of bound states. They can be seen as a core nucleus (which may be in an excited state) 
surrounded by one or two nucleon(s). The recent development of radioactive beams \cite{BNV13}
provides helpful information about the structure of exotic nuclei. The analysis of these 
data require models for the structure of the nucleus, as well as for the reaction process.

Over the last 20 years, the $\carb$ nucleus has been investigated in several works. From a Coulomb 
breakup experiment, Datta Pramanik $\etal$ \cite{DAB03} concluded that the ground state has
a spin $3/2^+$ and present a $\cartn$ structure. The study of excited states was performed by 
Elekes $\etal$ \cite{EDK05} by inelastic scattering on a proton target, and by Bohlen $\etal$
\cite{BHV07} who used a three-neutron transfer reaction at high energies.  Negative-parity states
were observed from $\beta$-delayed neutron emission of $^{17}$B \cite{UMY13}. Satou $\etal$ \cite{SNK08} 
observed several unbound states from proton inelastic scattering, and concluded on the existence of 
narrow $7/2^+$ and $9/2^+$ resonances are low energies.  The lifetime of excited states was 
investigated by Smalley $\etal$ \cite{SIN15} from a one-neutron knockout reaction on $\carc$. 
Very recently, this technique was also used by Kim $\etal$ \cite{KHS23} who concluded 
on the existence of a $5/2^+_2$ state, and suggested low-energy negative parity resonances. 

On the theoretical side, various techniques have been used to describe the spectroscopy of 
$\carb$, in particular the multi channel algebraic scattering (MCAS) model \cite{ACF12}, the no-core 
shell model \cite{SIN15} or large-scale shell model calculations \cite{KHS23}. In Refs.\ \cite{De00b,TD10},
the $\carb$ nucleus was described in the Resonating Group Method (RGM, see Refs.\ \cite{WT77,Ho77}). 
In that approach, the $\carb$ wave functions are obtained from a microscopic Hamiltonian, with 
the cluster approximation. In this way, the 17-nucleon antisymmetrization is exactly taken into account.  
This technique is well adapted to weakly bound nuclei, where the description of the relative 
wave function at large distances is a fundamental issue. The RGM provides the overlap integrals 
and the spectroscopic factors, which are necessary ingredients to the $(d,p)$ transfer cross sections
\cite{BT92}. 
No further parameter or renormalization factor is necessary, in contrast with the 
Distorted Wave Born Approximation (DWBA) 
method \cite{Sa83}, where the overlap integral is approximated from the simple potential model, and
where the spectroscopic factor is an adjustable parameter.  

An efficient tool to investigate the spin of nuclear states is provided by $(d,p)$ reactions, where a 
neutron is transferred from the deuteron to the target \cite{TJ20}. The cross sections are known to be very 
sensitive to the target+neutron angular momentum \cite{Sa83}. This technique was used recently by 
Pereira-L\'opez $\etal$ in a $\cdp$ experiment \cite{PFD20}. The authors measured the transfer cross sections 
to the $\carb$ $1/2^+$ and $5/2^+$ excited states, as well as the sum of the three bound states. 
Owing to its dominant $\cartn$ structure, the $\carb$ ground state presents a small spectroscopic 
factor in the $\carzn$ channel, and therefore the corresponding cross section could not be separated.

In the present work, our goal is to analyze the $\cdp$ reaction by using microscopic 
$\carb$ wave functions. In the standard DWBA treatment of $(d,p)$ reactions \cite{Sa83},
the residual nucleus is described by a potential-model wave function, which is renormalized by a 
spectroscopic factor to include missing effects, such as the antisymmetrization or the influence of 
core excited states.  This semi-microscopic approach to transfer reactions
has been developed in Ref.\ \cite{De22}, and we refer to this reference for more detail.

As the deuteron projectile is weakly bound, breakup effects in the 
$\card$ scattering wave functions are expected to be important. This problem  is addressed by 
using the Continuum Discretized Coupled Channel (CDCC) method (see Ref.\ \cite{YMM12} for a review), 
where the three-body continuum is simulated by pseudostates in the $p+n$ system. This technique is
well known, and has been used for many systems. It is particularly well adapted to exotic nuclei, 
where the breakup threshold energy is low, and where breakup effects are expected to be important 
(see, for example, Refs.\ \cite{GM17,De17}). 

The paper is organized as follows. In Section \ref{sec2}, we present a brief outline of the model, 
which is divided in two parts. In the first part, we discuss the RGM $\carb$ wave functions, in particular
the overlap integrals and the spectroscopic factors. The second part is focused on the $\card$ scattering 
wave functions, defined in the CDCC framework. We also provide information on the calculation of 
the transfer cross sections. The results on the $\carb$ spectroscopy and on the $\cdp$ cross sections 
are presented in Sec.\ \ref{sec3}. Conclusions and outlook are discussed in Sec.\ \ref{sec4}. 

\section{The model}
\label{sec2}
\subsection{The Resonating Group Method}
Our goal is to describe the $\carb$ wave functions in a microscopic approach.  In a $A$-nucleon system 
($A=17$), the Hamiltonian is given by
\begin{align}
	H_{17}=\sum_{i=1}^A t_i + \sum_{i<j=1}^A v_{ij},
	\label{eq1}
\end{align}
where $t_i$ is the kinetic energy of nucleon $i$, and $v_{ij}$ is a nucleon-nucleon interaction, 
including central and spin-orbit components.  The Coulomb force is treated exactly.
In the Resonating Group Method, the wave functions are defined at the cluster approximation and 
involve internal wave functions of the clusters, as well as a relative function.  The internal 
wave functions are defined in the shell model.  For the $\carb$ nucleus, the RGM wave function
associated with Hamiltonian $H_{17}$  reads
\begin{align}
	\Psi_{17}^{JM\pi}={\cal A}\frac{1}{\rho}\sum_{c} \varphi^{JM\pi}_{c}g^{J\pi}_{c}(\rho) , 
	\label{eq2}
\end{align}
where ${\cal A}$ is the $A$-nucleon antisymmetrizor, $\pmb{\rho}=(\rho,\Omega_{\rho})$ is the 
relative coordinate between $\cara$ and the neutron, and $g^{J\pi}_{c}(\rho)$ are the radial wave functions.
The channel functions $\varphi^{JM\pi}_{c}$ are given by
\begin{align}
\varphi^{JM\pi}_{c}=\biggl[\bigl[\phi^{I_1}_{16} \otimes \phi_{n}\bigr]^{I} \otimes Y_{\ell}(\Omega_{\rho})\biggr]^{JM},
\label{eq3}
\end{align}
where $\phi^{I_1}_{16}$ are shell-model wave functions of $\cara$ (with an oscillator 
parameter $b$, chosen here as $b=1.6$ fm), $\phi_n$ is a neutron spinor, and index $c$ 
stands for $c=(\ell,I_1,I)$.  We adopt the same conditions as in Ref.\ \cite{TD10}, where $\cara$ is described 
by all Slater determinants involving four protons in the $p$ shell, and two neutrons in the $sd$  shell.  
This leads to 990 Slater determinants to describe the ground state and several excited 
states (see Ref.\ \cite{TD10} for detail).

In the calculation of transfer cross sections, the relevant quantities are the overlap integrals
\cite{Ti14,Ti20}, defined as
\begin{align}
	I^{J\pi}_{c}(\rho)=\frac{1}{\rho}\bigl\langle \varphi^{JM\pi}_{c} \vert \Psi_{17}^{JM\pi} \bigr\rangle,
	\label{eq6}
\end{align}
which are defined for each channel $c$.  This definition provides the spectroscopic factor from
\begin{align}
	S^{J\pi}_{c}=\int	\left[ I^{J\pi}_{c}(\rho)\right]^2  d\rho .
	\label{eq7}
\end{align}

The relative functions $g^{J\pi}_{c}(\rho)$ must be determined from the Schr\"odinger equation 
associated with Hamiltonian (\ref{eq1}).  In practice, we use the Generator Coordinate Method (GCM,
see Refs.\ \cite{Ho77,DD12}), where $g^{J\pi}_{c}(\rho)$ is expanded as
\begin{align}
	g^{J\pi}_{c}(\rho)=\int dR \, f^{J\pi}_{c}(R)\, \Gamma_{\ell}(\rho,R).
\label{eq4}
\end{align}
In this equation, $R$ is the generator coordinate, and the projected Gaussian function 
$\Gamma_{\ell}(\rho,R)$ is defined as
\begin{align}
	\Gamma_{\ell}(\rho,R)=\left(\frac{\mu}{\pi b^2}\right)^{3/2}
\exp\left(-\frac{\mu}{2b^2}(\rho^2+R^2)\right) i_{\ell}\left(\frac{\mu \rho R}{b^2}\right),
	\label{eq5}
\end{align}
$\mu$ being the reduced mass and $i_{\ell}(x)$ a spherical Hankel function.  
With expansion (\ref{eq4}), the total wave function (\ref{eq2}) can be written as a superposition of 
projected Slater determinants \cite{DD12}, well adapted to systematic numerical calculations. 
In the GCM, the calculation of the radial functions $g^{J\pi}_{c}(\rho)$ is therefore replaced by the calculation of
the generator functions $f^{J\pi}_{c}(R)$. 

The Gaussian asymptotic behaviour (\ref{eq5}) is corrected with the microscopic $R$-matrix 
method \cite{DB10} for scattering states as well as for bound states.  This issue is important 
for weakly bound states, where the wave function presents a slow decrease at large distances.
The technique to derive the overlap integrals in the RGM is explained, for example, in Refs.\ \cite{De22,De23}.  
It is based on the calculation of the overlap kernels between GCM basis functions.

The Asymptotic Normalization Coefficient (ANC) $C^{J\pi}_{c}$ in channel $c$ is defined 
from the long-range limit of the overlap integral as
\begin{align}
	I^{J\pi}_{c}(\rho)\longrightarrow C^{J\pi}_{c} W_{-\eta_c,\ell+1/2}(2k_c\rho),
	\label{eq17}
\end{align}
where $k_c$ and $\eta_c$ are the wave number and Sommerfeld parameter in channel $c$, and $W_{ab}(x)$ 
is the Whittaker function \cite{NIST}.  The ANC depends on the wave function at 
large distances, whereas the spectroscopic factor probes the inner part of the wave function.  
Both quantities are therefore complementary. Notice that the asymptotic forms of the relative wave
functions $g^{J\pi}_{c}(\rho)$ and of the overlap integrals $I^{J\pi}_{c}(\rho)$ are identical.

\subsection{$\card$ and $\carp$ scattering wave functions}
In the entrance channel, the $\card$ scattering wave functions are defined in the CDCC formalism, 
which simulates breakup effects of the deuteron by $p+n$ pseudostates \cite{YMM12}.  
The three-body Hamiltonian is defined as
\begin{align}
	H=H_0(\pmb{r})+T_R + V_{pC} +V_{nC},
	\label{eq8}
\end{align}
where $H_0$ is the $p+n$ Hamiltonian, $T_R$ the $\card$ kinetic energy, and $V_{pC}$ and $V_{nC}$ are 
optical potentials between the nucleons and $\cara$. In Eq.\ (\ref{eq8}), $\pmb{r}$ is the $p+n$ 
relative coordinate, and $\pmb{R}$ is associated with the $\card$ system.

In the CDCC method, $p+n$ wave functions are obtained from
\begin{align}
	H_0 \phi^{\ell m}_k(\pmb{r})=E^{\ell}_k \phi^{\ell m}_k(\pmb{r}),
	\label{eq9}
\end{align}
where $k$ is the excitation level, and $\ell$ the angular momentum.  This equation provides one 
bound state $(E^0_1 < 0)$, associated with the deuteron, and positive-energy states $(E^{\ell}_k > 0)$,
referred to as pseudostates, which represent approximations of the continuum.  The $\card$ wave 
functions are then defined as
\begin{align}
	\Psi^{JM\pi}_{i}=\frac{1}{R}\sum_{\gamma}u^{J\pi}_{\gamma}(R)
	\left[ \phi^{\ell}_k(\pmb{r}) \otimes Y_L(\Omega_R)\right] ^{JM},
	\label{eq10}
\end{align}
where $L$ is the relative orbital momentum between $\cara$ and $d$, and where index $\gamma$ stands for $\gamma=(\ell,k,L)$.  
The summation over the pseudostates $(\ell,k)$ must be truncated at some $\lmax$ and $\emax$ values, 
which are chosen large enough so that the expansion (\ref{eq10}), and the associated cross sections converge.

The radial functions $u^{J\pi}_{\gamma}(R)$ and the associated scattering matrices at a center-of-mass 
(c.m.) energy
$E$ are obtained 
from the standard coupled-channel system
\begin{align}
	\bigl( T_{L}(R)+E^{\ell}_k-E\bigr) u^{J\pi}_{\gamma}(R)+
	\sum_{\gamma'}V^{J\pi}_{\gamma, \gamma'}(R) u^{J\pi}_{\gamma'}(R)=0,
	\label{eq11}
\end{align}
with
\begin{align}
	T_L(R)=-\frac{\hbar^2}{2\mu}\biggl(\frac{d^2}{dR^2}-\frac{L(L+1)}{R^2}\biggr).
	\label{eq12}
\end{align}
The coupling potentials $V^{J\pi}_{\gamma,\gamma'}(R)$ are obtained from matrix elements of the 
optical potentials $V_{pC}+V_{nC}$ \cite{YMM12}.  Finally, system (\ref{eq11}) is solved with the $R$-matrix 
method \cite{DB10,Bu11} on a Lagrange mesh \cite{De16a} which provides the scattering matrices and the radial wave functions
for all $J\pi$ values.

For the $\carp$ exit channel, the scattering wave functions are defined from an optical potential as
\begin{align}
	\Psi^{JM\pi}_f=\frac{1}{R'}u^{J\pi}_{f}(R')
	\left[ \bigl[\Psi_{17}\otimes \phi_p \bigr]^{I_f} \otimes Y_{L_f}(\Omega_{R'})\right]^{JM},
	\label{eq13}
\end{align}
where $\pmb{R}'$ is the $\carp$ relative coordinate, and where $L_f$ and $I_f$ are the angular momentum
and the channel spin in the exit channel. Function $u^{J\pi}_{f}(R')$ is obtained from the Schr\"odinger equation
\begin{align}
	\bigl( T_{L_f}(R')+V_{\rm opt}(R')-E_f\bigr) u^{J\pi}_{f}(R')=0
	\label{eq13b}
\end{align}
which involves the optical potential $V_{\rm opt}(R')$, associated with $\carp$ ($E_f$ is the scattering energy in this channel).

\subsection{Transfer scattering matrices}
We follow the method of Ref.\ \cite{De22}, and refer the reader to this reference for detail.  
The $(d,p)$ transfer scattering matrix from an initial state $i$ to a final state 
$f$ is defined by
\begin{align}
	U^{J\pi}_{i,f}=-\frac{i}{\hbar}
	\langle \Psi^{J\pi}_{f} \vert V_{pn}+\Delta V \vert  \Psi_{i}^{J\pi}\rangle,	
	\label{eq14}
\end{align}
where we use the post representation, and where $\Delta V$ is the remnant potential \cite{Sa83,TJ20}.  A 
difficulty associated with definition (\ref{eq14}) is that the coordinates $(\pmb{R},\pmb{r})$ in 
the entrance channel are different from those in the exit channel  $(\pmb{R}',\pmb{r}')$.  
In practice, $\pmb{r}$ and $\pmb{r}'$ are expressed from $\pmb{R}$ and $\pmb{R}'$ 
(see, for example, Ref.\ \cite{SD19} for detail).  Using definitions (\ref{eq10}) and (\ref{eq13}) 
of the scattering wave functions, the transfer scattering matrix (\ref{eq14}) can be reformulated as
\begin{align}
	U^{J\pi}_{i,f}=-\frac{i}{\hbar}\sum_{\gamma} \int u^{J\pi}_{\gamma}(R) K^{J\pi}_{\gamma}(R,R')
	u^{J\pi}_{f}(R') R R'dR dR'.
	\label{eq15}
\end{align}
For a given $\carb$ state, the transfer kernel is given by
\begin{align}
	&	K^{J\pi}_{\gamma}(R,R')={\cal J}  \sum_c \nonumber \\
	&	\langle \bigl[ \phi_k^{\ell}(\pmb{r}) \otimes Y_{L_i}(\Omega_R)\bigr]^{J} \vert V_{pn}+\Delta V \vert
	\bigl[ I_{c}(\pmb{r}') \otimes Y_{L_f}(\Omega_{R'})\bigr]^{J} \rangle,
	\label{eq16}
\end{align}
where ${\cal J}$ is the Jacobian, and $I_{c}(\pmb{r}')$ are the overlap functions defined in 
Eq.\ (\ref{eq6}). In Eq.\ (\ref{eq15}), the summation over index $\gamma$ arises from the 
CDCC expansion of the $\card$ 
scattering wave functions.  The use of a Lagrange mesh to define the radial functions 
$u^{J\pi}_{\gamma}(R)$ makes the numerical calculations of the double integrals (\ref{eq15})
rather fast \cite{SD19}. The $R$-matrix channel radius is chosen large enough to guarantee the 
convergence.

\section{The $\cdp$ transfer reaction}
\label{sec3}

\subsection{Overlap integrals of $\carb$}
The conditions of the calculations are as in Ref.\ \cite{TD10}, where a multichannel $\carn$ RGM calculation 
is performed with all $(\pi\, p)^4(\nu\, sd)^2$ $\cara$ configurations (990 Slater determinants).  
The spectra of $\cara$ and $\carb$ given 
in Ref.\ \cite{TD10} agree reasonably well with experiment.  We use the Volkov $V2$ 
interaction, complemented by a zero-range spin-orbit force.  The Majorana parameter is $M=0.668$, and 
the spin-orbit amplitude is $31.3$ MeV.fm$^5$.  With these parameters, $3/2^+$, $1/2^+$ and $5/2^+$ 
states are obtained below the $\carn$ threshold, in agreement with experiment.  For the calculation of 
the overlap integrals and of the spectroscopic factors, a slight readjustment of $M$ is introduced in 
order to reproduce exactly the experimental binding energies ($-0.734,\ -0.517$ and $-0.402$ 
MeV, respectively).

A useful information on the structure of a nucleus is provided by the rms radius, defined in a translation-invariant form as
\begin{align}
\langle r^2 \rangle=\frac{1}{A}\langle \Psi^{J\pi}_{17} \vert
\sum_{i=1}^A (\pmb{r}_i-\pmb{R}_{\rm c.m.})^2 
\vert \Psi^{J\pi}_{17}\rangle,
	\label{eq16b} 
\end{align}
where $\pmb{R}_{\rm c.m.}$ is the center-of-mass of the system. The rms radius of the $\cara$
core is given by the shell-model value $\sqrt{\langle r^2 \rangle}=\sqrt{73/32}b=2.42$ fm with $b=1.6$ fm (see Sec. \ref{sec2}.A).

\begin{figure}[htb]
	\centering
	\includegraphics[width=7.6cm]{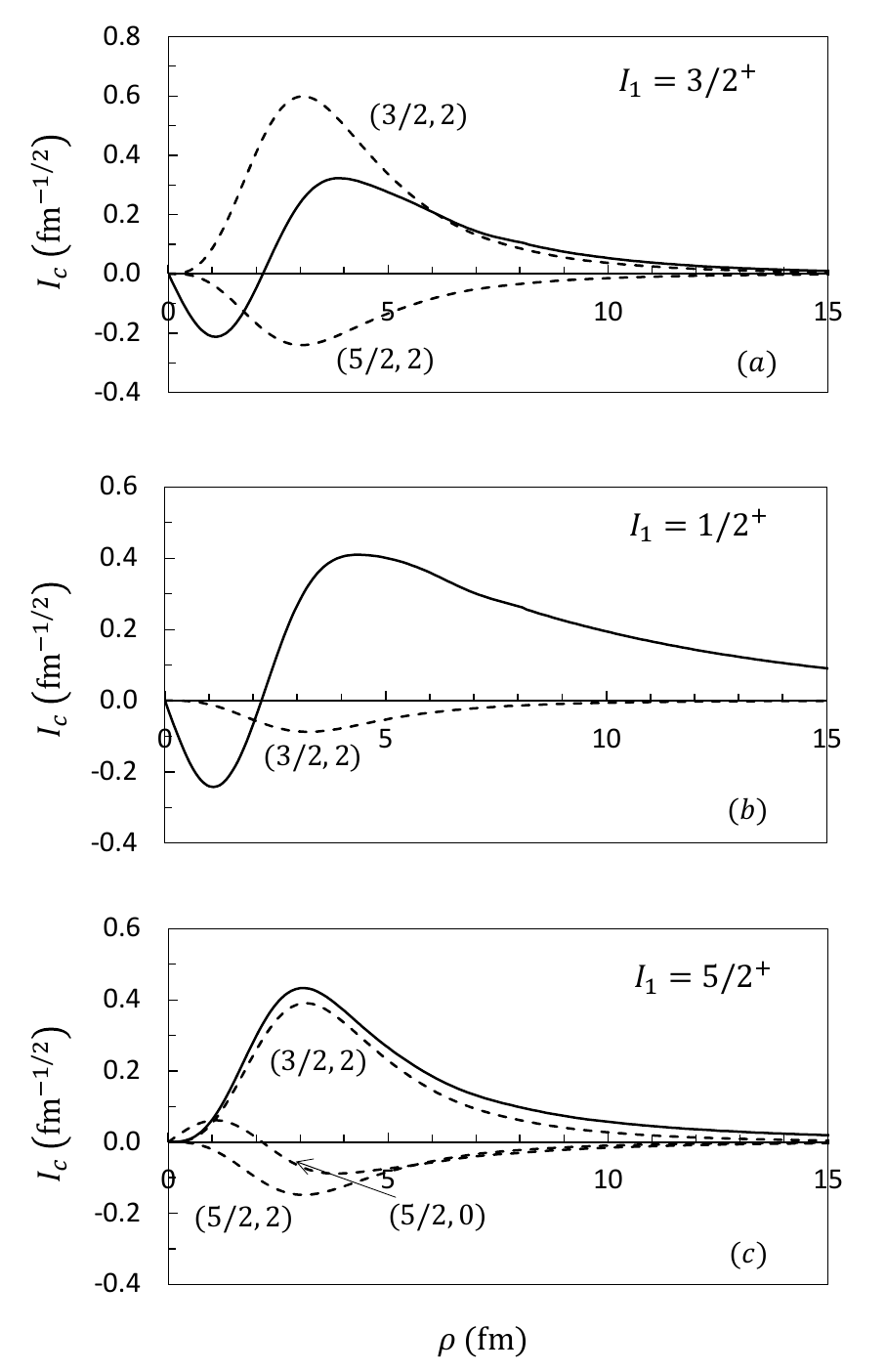}
	\caption{Overlap integrals of the $\carb$ bound states with spin $I_1$. The solid and dashed lines correspond to the
	$\carzn$ and $\cartn$ configurations, respectively. The labels refer to the $(I\ell)$ values (see
also Table \ref{table1}).}
	\label{fig_oi}
\end{figure}

In Fig.\ \ref{fig_oi}, we show the overlap integrals for the three bound states, and we provide 
the corresponding spectroscopic factors and ANC in Table \ref{table1}.  
Let us first discuss the $3/2^+$ ground state shown in Fig.\ \ref{fig_oi}(a).  This $\carb$ state is 
essentially described by a $\cartn$ configuration.  The overlap integral in the $\carzn$ channel is 
quite small.  The main contribution to the spectroscopic factor comes from the $(I,\ell)=(3/2,2)$ 
channel with $S=0.97$.  As mentioned in Ref.\ \cite{PFD20}, the $\cdp$ cross section to the ground state 
is expected to be small, owing to the small spectroscopic factor in this channel  ($S=5.2\times 10 ^{-3}$).
The ground-state radius is in reasonable agreement with a recent experiment ($2.68\pm 0.05$ fm) \cite{DKT21},
where interaction cross sections are analyzed in the Glauber model.

\begin{table}[htb]
	\caption{Spectroscopic factors $S$, ANC (in fm$^{-1/2}$) and rms radii $\rms$ (in fm) of the $\carb$ bound
		states (in the first column, the energy is given in MeV). The column $(I\ell)$ refers to the various components in the $\cartn$ channel.
		The notation $x^y$ stands for $x\times 10^y$.
	\label{table1}}
\begin{ruledtabular}
	\begin{tabular}{c|rrcrrc}
		$I_1$ & $S(0^+)$ & $C(0^+)$ & $(I,\ell)$ & $S(2^+)$ & $C(2^+)$ & $\rms$\\
	\hline
		$3/2^+$ & $5.2^{-3}$ & $7.4^{-3}$ & $(3/2,0)$ & $   0.370$ & $   1.596$ & 2.61\\
		$(-0.734)$ &  &  & $(3/2,2)$ & $   0.972$ & $   0.517$\\
		&  &  & $(5/2,2)$ & $   0.156$ & $-  0.205$\\
		&  &  &  &  & \\
		$1/2^+$ & $   0.942$ & $   0.959$ & $(3/2,2)$ & $   0.021$ & $-0.080$ & 2.75\\
	$(-0.517)$ 	&  &  & $(5/2,2)$ & $1.1^{-3}$ & $-  0.026$\\
		&  &  &  &  & \\
		$5/2^+ $ & $   0.562$ & $   0.045$ & $(3/2,2)$ & $   0.423$ & $   0.290$ & 2.60\\
		$(-0.402)$ &  &  & $(5/2,0)$ & $   0.029$ & $-  0.380$\\
		&  &  & $(5/2,2)$ & $   0.060$ & $-  0.106$\\
	\end{tabular}
\end{ruledtabular}
\end{table}

In contrast, the $1/2^+$ first excited state has a dominant $\carzn$ structure.  The low binding 
energy ($-0.517$ MeV) and the angular momentum $\ell=0$ suggest a halo structure, which is supported 
by the large ANC value and by the large rms radius. The GCM radius corresponds to a $\carn$ distance of
6.10 fm (in comparison, this distance is 4.78 fm for the ground state).  The $\cartn$ channel plays a minor role, essentially at short distances.  

The $5/2^+$ second excited state is more complex, since similar overlap integrals are obtained in 
the $\carzn$ and $\cartn$ channels.  Notice that the 
calculation involves many other $\carn$ channels, which have an impact on the $\carb$ wave functions 
(\ref{eq2}).  However, the corresponding spectroscopic factors are small, and therefore are not shown here.

\subsection{$\card$ elastic-scattering cross sections}
The calculation of the $\card$ scattering wave functions is the first step for the transfer cross sections.  
The elastic cross section also provides an excellent test of the model.  For $\card$, elastic cross 
sections have been measured at $\elab=24\ {\rm MeV/nucleon}$ ($\ecm=42.67$ MeV) in Ref.\ \cite{JLY20}, which 
is slightly higher than the energy of the $\cdp$ experiment (Ref.\ \cite{PFD20}, $\ecm=30.58$ MeV).  
However, this energy difference is not expected to modify the conclusions on the reliability of the model.

The calculation is performed within the CDCC framework, with the Minnesota potential \cite{TLT77}
for the $p+n$ interaction. We use a Laguerre basis to describe the 
deuteron ground state, as well as the $p+n$ pseudostates (see Ref.\ \cite{DBD10} for detail).
Pseudostates up to $\emax=20$ MeV and 
$\lmax=4$ are included.   The $\card$ scattering 
matrices are computed with the $R$-matrix formalism \cite{DB10,De16a}.  Several tests of the 
numerical conditions ($\emax$, $\lmax$, $R$-matrix radius, $\card$ basis, etc.) have been performed to 
check the stability of the cross section.  As $^{16}$C + nucleon optical potentials, we use the 
global parametrizations of Ref.\ \cite{KD03} and of Ref.\ \cite{VTM91}, referred to as KD and 
CH, respectively.  

\begin{figure}[htb]
\centering
\includegraphics[width=8.6cm]{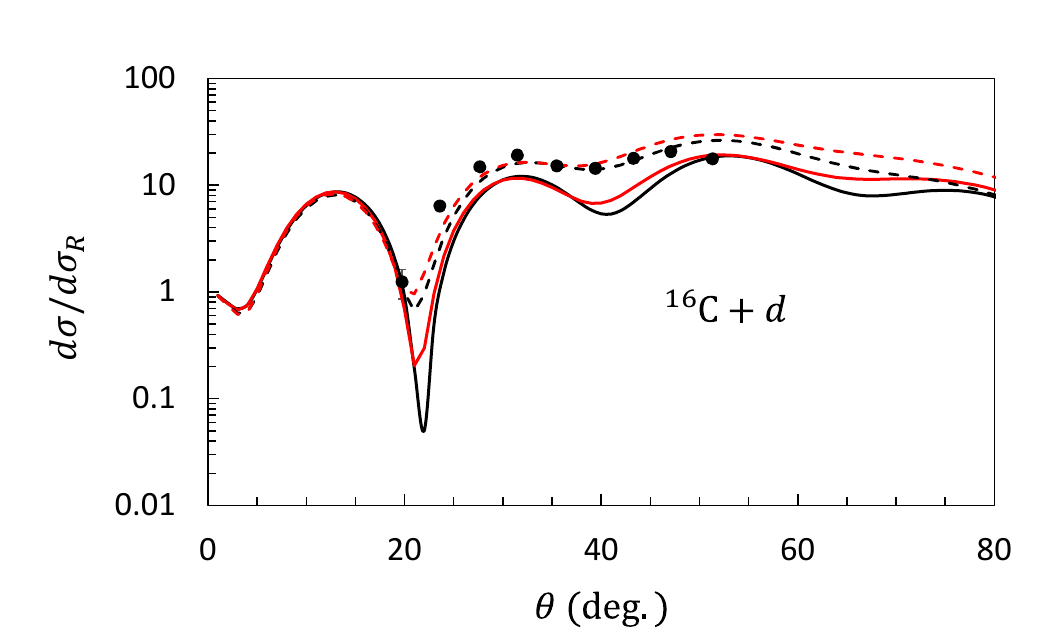}
\caption{Elastic $\card$ cross section (divided by the Rutherford cross section) with the CDCC wave functions (solid lines) and with the single-channel approximation (dashed lines).
The black and red lines
correspond to the KD and CH $^{16}$C+nucleon optical potentials, respectively.
Experimental data are taken from Ref.\ \cite{JLY20}}
\label{fig_ela}
\end{figure}

The $\card$ elastic cross section at $\ecm=42.67$ MeV is presented in Fig.\ \ref{fig_ela}, with 
the experimental data of Jiang $\etal$ \cite{JLY20}.  We show the full CDCC cross sections, as well
as the single-channel approximation, where breakup effects of the deuteron are neglected.
The minimum near $\theta=20^{\circ}$ is 
well reproduced, and the theoretical cross section is weakly dependent on the  $^{16}$C + nucleon optical potential.  The present calculation is consistent with results of Ref.\ \cite{SD22}, where a five-body 
$\card$ CDCC calculation was performed.  It was shown in that reference that breakup effects tend to 
reduce the cross section for $\theta \gtrsim 30^{\circ}$. This leads to an underestimation of the cross section,
although the shape is consistent with experiment. The availability of experimental data at other energies
would be helpful to clarify the role of breakup effects in $\card$ elastic scattering.

\subsection{$\cdp$ and $\cpd$ transfer cross sections}
\begin{figure}[htb]
\centering
\includegraphics[width=8.6cm]{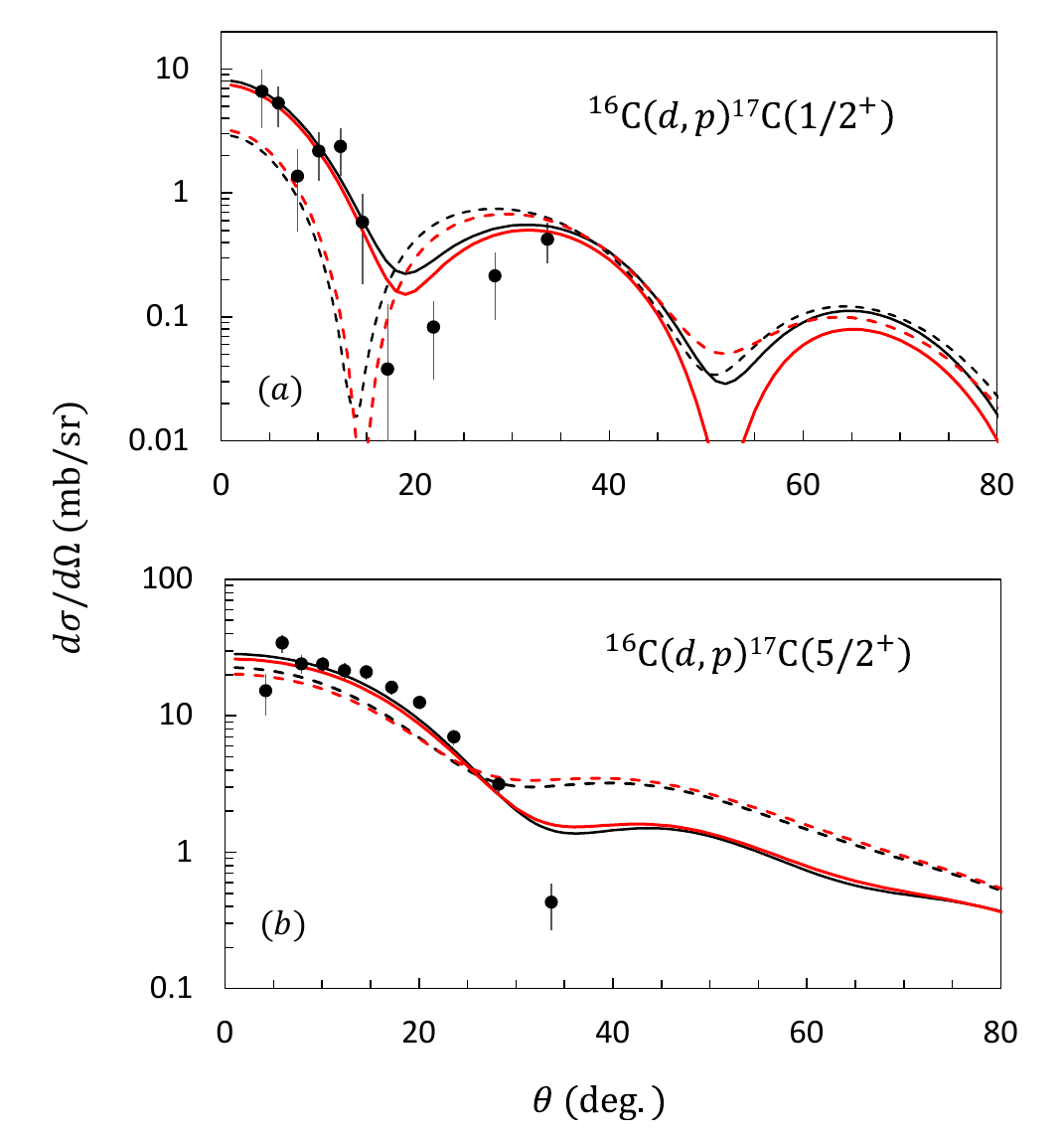}
\caption{$\cdp$ cross sections to the $1/2^+$ (a) and $5/2^+$ (b) states. The solid and dashed lines
correspond to the full CDCC calculation, and to the no-breakup approximation, respectively.
Results with the KD and CH $^{16}$C+nucleon optical potentials are shown in black and red, respectively. 
The experimental data are taken from Ref.\ \cite{PFD20}, and transformed from the lab frame to the c.m. frame.}
\label{fig_cdp_15}
\end{figure}

Figure \ref{fig_cdp_15} presents the $\cdp$ cross section to the $1/2^+$ and $5/2^+$ states, and compared to experiment \cite{PFD20}. 
In  Ref.\ \cite{PFD20}, the authors compute the cross sections with the Adiabatic Distorted Wave approximation, which involve some parameters. In contrast, the present calculations do not contain any adjustable parameter since the initial state 
is determined from the $^{16}$C+nucleon potentials, and the $\carb$ final states by the RGM 
overlap integrals. The spectroscopic factor is not a parameter, but an output of the microscopic model.
For both states, the cross sections are weakly dependent on the $\cara$+nucleon optical potential,
and the CDCC model is in nice agreement with experiment \cite{PFD20}, although 
$\cdp (1/2^+)$ cross section near the minimum at $\theta \approx 20^{\circ}$  is overestimated. The single-channel approximation is obviously less good than the full CDCC approach. The quality 
of the theoretical cross sections, particularly at small angles, supports the spectroscopic factors 
presented in Table \ref{table1}. 

\begin{figure}[htb]
\centering
\includegraphics[width=8.6cm]{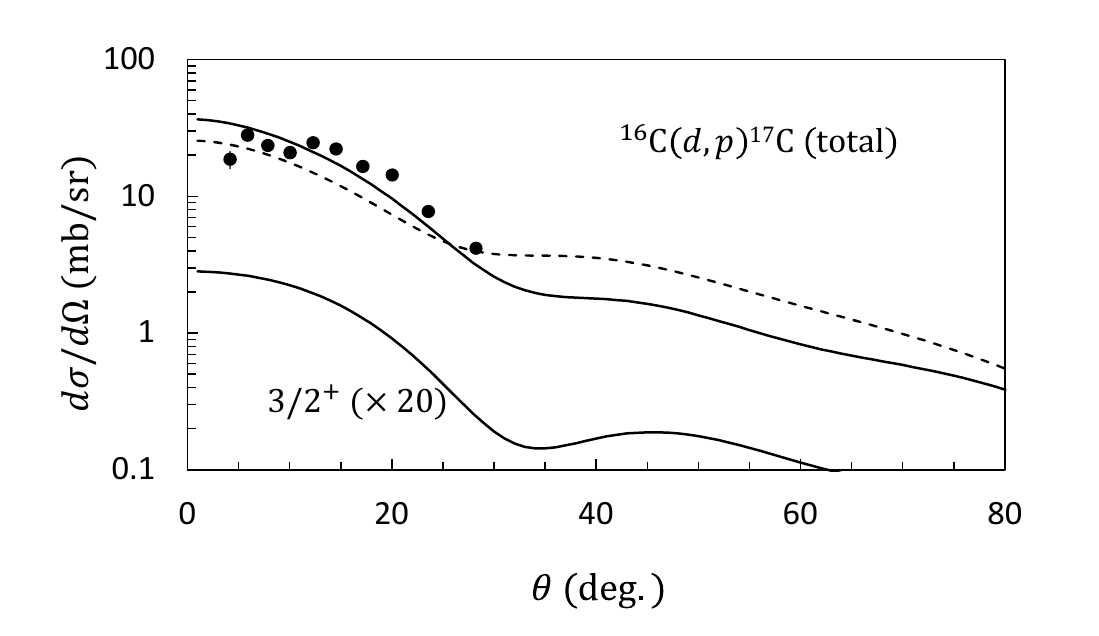}
\caption{Ground-state ($\times 20$) and total $\cdp$ cross sections with the KD potential 
(the CH results are very similar). See caption to Fig.\ \ref{fig_cdp_15}.}
\label{fig_cdp_tot}
\end{figure}

The contribution of the $3/2^+$ ground state, shown in Fig.\ \ref{fig_cdp_tot}, 
deserves a special attention. Since this state 
has a dominant $\cartn$ structure, a standard DWBA approach, using a $\carn$ potential of $\carb$, cannot be 
used. The spectroscopic factor aims at correcting the normalization of a potential-model wave function.
However, if the spectroscopic factor is very small ($S=5.2\times 10^{-3}$ in the present case), 
the shape of the approximated wave function is questionable. Consequently, a microscopic (multichannel)
approach is well appropriated for the $\carb$ ground state.
The cross section is expected to be small (see Fig.\ \ref{fig_cdp_tot}) and could not be separated in 
the experiment of Ref.\ \cite{PFD20}. Figure \ref{fig_cdp_tot} suggests that the cross section 
around $\theta=0$ is of the order of 0.1 mb, to be compared with 10-20 mb for the excited 
states. Even if these states contribute very little to the total cross section, a specific measurement
would be helpful to assess the validity of the overlap integral provided by the RGM. 

Figure \ref{fig_cdp_tot} also contains the total cross section,  where the contribution of the ground 
and excited $\carb$ states are summed. As mentioned above, the full CDCC calculation agrees reasonably 
well with experiment, whereas the single-channel approximation underestimates the data for $\theta \gtrsim 10^{\circ}$. 

\begin{figure}[htb]
	\centering
	\includegraphics[width=8.6cm]{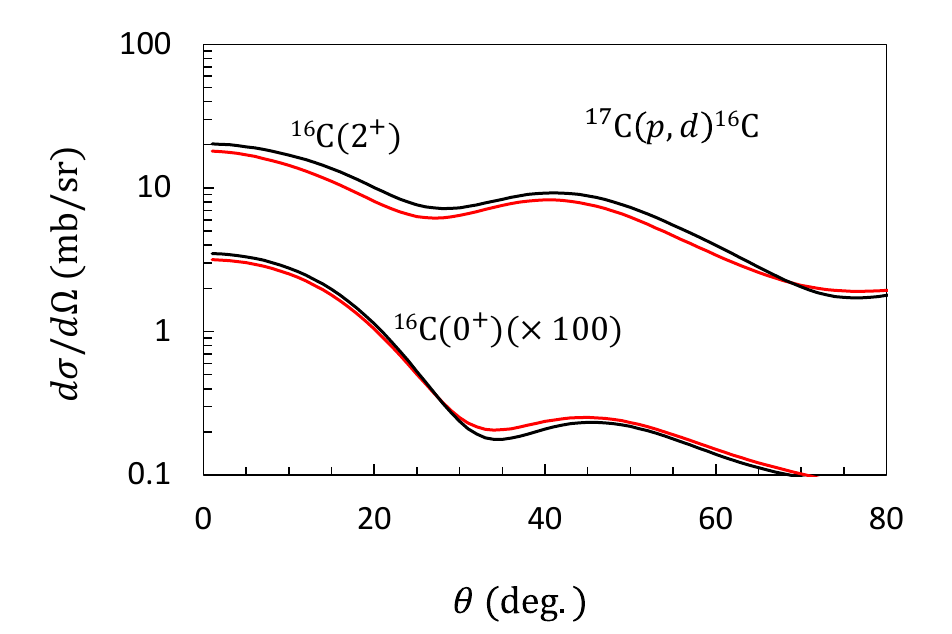}
	\caption{$\cpd$ cross sections to the $0^+$ and $2^+$ states of $\cara$ at $\ecm=29.1$ MeV (see text). 
		The black and red lines correspond to the KD and CH $^{16}$C+nucleon optical potentials.}
	\label{fig_cpd}
\end{figure}

As the $\cdp$ cross section to the ground state is small and difficult to measure, we consider 
the reverse $\cpd$ reaction. The corresponding cross sections are obtained as explained in Sec.\
\ref{sec2}.C. In practice, we compute the $^{16}{\rm C}(2^+)(d,p)^{17}$C cross section, 
and use the detailed balance theorem to deduce the $^{17}{\rm C}(p,d)^{16}{\rm C}(2^+)$ values.
The $\carp$ relative energy is therefore slightly shifted by the $Q$ value ($\ecm=29.1$ MeV).
The  cross sections to the $0^+$ and $2^+$ states are shown in Fig.\ \ref{fig_cpd}. 
The calculation of the $^{16}{\rm C}(2^+)(d,p)^{17}$C cross section is similar to the previous
$^{16}{\rm C}(0^+)(d,p)^{17}$C calculation, but is sensitive to the $\cartn$ overlap integral of $\carb$. The
nucleon+$^{16}$C optical potentials are unchanged.

The $2^+$ 
contribution is of the order of 10 mb at small angles, similar to the values found for $\cdp$. 
As expected, the $0^+$ is small, and lower and 0.1 mb. However, 
such a measurement would be a stringent test of the RGM, as it would probe a small component of the $\carb$
wave function, more sensitive to coupling effects. As for the $\cdp$ cross sections, the sensitivity to the optical potential is low.  

\section{Conclusion}
\label{sec4}
We have analyzed recent data on the $\cdp$ reaction \cite{PFD20} by using microscopic $\carb$ wave 
functions. In the microscopic RGM formalism, the wave functions are fully antisymmetric. The 
$\carn$ cluster structure includes many $\cara$ states, which provides a realistic description of 
the low-lying states. The long-range behavior is treated by the $R$-matrix methods, for bound states 
as well as for scattering states. The $\card$ wave functions are obtained within the 
CDCC approach, which includes deuteron breakup effects. We have shown that these effects are not
negligible in the transfer cross sections. 
 
The transfer cross sections are in fair agreements with experiment, without any adjustable parameter. 
We have confirmed that the $\cdp({\rm gs})$ cross section is small, owing to the dominant $\cartn$ 
structure of the $\carb$ ground state. We have determined the reverse $\cpd$ cross section, 
and shown that $\cara$ would be essentially populated in the $2^+$ first excited state. A 
measurement of the cross sections, and in particular of the branching ratio, should provide a 
strong test of the RGM wave function. 
 
\section*{Acknowledgments}
We are grateful to B. Fern\'andez-Dom\'inguez for providing us with the cross section data of Ref.\ \cite{PFD20}.
This work was supported by the Fonds de la Recherche Scientifique - FNRS under Grant Numbers 4.45.10.08 and J.0065.22.
	
%

\end{document}